\documentclass[aps,prb,twocolumn,superscriptaddress]{revtex4}

\usepackage{graphicx}

\begin{document}


\title{ Suppression of rectification at metal-Mott-insulator interfaces }


\author{ Kenji Yonemitsu }
\email[]{kxy@ims.ac.jp}
\affiliation{ Institute for Molecular Science, Okazaki 444-8585, Japan }
\affiliation{ Department of Functional Molecular Science, Graduate University for Advanced Studies, Okazaki 444-8585, Japan }
\author{ Nobuya Maeshima }
\affiliation{ Institute for Molecular Science, Okazaki 444-8585, Japan }
\affiliation{ Department of Chemistry, Tohoku University, Aramaki, Aoba-ku, Sendai 980-8578, Japan }
\author{ Tatsuo Hasegawa }
\affiliation{ Correlated Electron Research Center (CERC), AIST, Tsukuba 305-8562, Japan }

\date{\today}

\begin{abstract}
Charge transport through metal-Mott-insulator interfaces is studied and compared with that through metal-band-insulator interfaces. For band insulators, rectification has been known to occur owing to a Schottky barrier, which is produced by the work-function difference. For Mott insulators, however, qualitatively different current-voltage characteristics are obtained. Theoretically, we use the one-dimensional Hubbard model for a Mott insulator and attach to it the tight-binding model for metallic electrodes. A Schottky barrier is introduced by a solution to the Poisson equation with a simplified density-potential relation. The current density is calculated by solving the time-dependent Schr\"odinger equation. We mainly use the time-dependent Hartree-Fock approximation, and also use exact many-electron wave functions on small systems for comparison. Rectification is found to be strongly suppressed even for large work-function differences. We show its close relationship with the fact that field-effect injections into one-dimensional Mott insulators are ambipolar. Experimentally, we fabricated asymmetric contacts on top of single crystals of quasi-one-dimensional organic Mott and band insulators. Rectification is strongly suppressed at an interface between metallic magnesium and Mott-insulating (BEDT-TTF)(F$_2$TCNQ) [BEDT-TTF=bis(ethylenedithio)tetrathiafulvalene, F$_2$TCNQ=2,5-difluorotetracyanoquinodimethane].
\end{abstract}

\pacs{73.40.Rw, 73.40.Ei, 73.20.Mf, 72.20.-i}
\keywords{metal-insulator interface, rectification}

\maketitle

\section{Introduction}
When a new function of some material like an organic molecular crystal is sought, one often considers a possible electronic device made of the material. Around the device, there always exists an interface between two materials with different transport properties and different work functions. Charge transport through metal-insulator or metal-semiconductor interfaces has been studied from long ago. \cite{schottky_38,mott_38} However, most of the theories are restricted to metal-band-insulator interfaces. \cite{sze-ng_book07} Now it has become clarified that many organic conductors and insulators are strongly correlated electron systems. \cite{jpsj_organic06} Their insulating phases cannot be explained on the basis of band structures. Electron-electron interactions play an essential role in the insulating phases of the (BEDT-TTF)$_2X$ salts. Then, conventional theories cannot be applied to charge transport through metal-Mott-insulator interfaces.

As is well known, two different materials generally have different work functions. When they are attached to each other, the energy levels are modified around the interface by transferring electrons to match their Fermi levels. This is often called band bending, and forms a Schottky barrier at the metal-insulator interface. The electronic state in the close vicinity of the interface can be modified physically and/or chemically and depends sensitively on the two materials. \cite{ishii_am99} Nevertheless, the overall band structure is governed by the long-range Coulomb interaction, i.e., by the Poisson equation if the band structure can be continuously treated. \cite{sze-ng_book07}

When a Schottky barrier dominates the characteristics of a field-effect transistor fabricated on an organic insulator, the field-effect characteristics indeed depend on whether it is a band insulator or a Mott insulator. For instance, when such transistors are fabricated on semiconducting carbon nanotubes, which are band insulators, the field-effect characteristics are generally unipolar unless the work function of the metallic electrodes is matched with that of the carbon nanotube. \cite{heinze_prl02,appen_prl02,rado_apl03,javey_n03} Now, fine control of $ p $-, $ n $-, and ambipolar-type operations is available in single-crystal organic field-effect transistors with the use of chemically tunable Fermi energies in tetrathiafulvalene-tetracyanoquinodimethane-based organic metal electrodes. \cite{takahashi_apl06} 
On the other hand, when metal-insulator-semiconductor field-effect transistor device structures are based on organic crystals of the quasi-one-dimensional Mott insulator (BEDT-TTF)(F$_2$TCNQ), the field-effect characteristics are ambipolar even if its work function is quite different from that of the metallic electrodes. \cite{hasegawa_prb04} We have theoretically shown that backward scatterings at interfaces and umklapp scatterings inside the Mott insulator are balanced, leading to collective charge transport insensitive to the band bending. \cite{yonemitsu_jpsj05}

Then, how about a simpler structure of just a metal-insulator interface without gate electrode? In order for the current-voltage characteristics to be observed, two different metallic electrodes need to be attached to an insulator in such a way that only one interface has a large work-function difference. The magnitude of the current density depends on the sign of the applied voltage: the forward (reverse) voltage lowers (raises) the Schottky barrier at this interface, leading to a larger (smaller) current density for band insulators including ordinary semiconductors. \cite{sze-ng_book07} If this asymmetry is large, it can be regarded as rectification. Such rectifying nature is observed in single-crystal organic devices without gate dielectric layers. \cite{takahashi_apl06} 
Then, what happens to the current-voltage characteristics if they are replaced by Mott insulators? In this paper, we show that the rectification is strongly suppressed for Mott insulators. Preliminary theoretical results are presented in Ref.~\onlinecite{yonemitsu_pacifichem}.

\section{One-Dimensional Models for Metal-Insulator Interfaces \label{sec:model}}

\begin{figure}
\includegraphics[height=1.5cm]{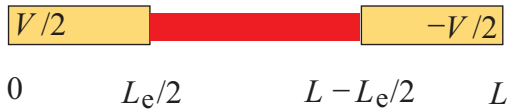}
\caption{(Color online) One-dimensional model for an insulator, to which two metallic electrodes are attached. 
\label{fig:geometry}}
\end{figure}

We use the one-dimensional Hubbard model for a Mott insulator (the one-dimensional tight-binding model with alternating transfer integrals for a band insulator) attached to two different metallic electrodes represented by the one-dimensional tight-binding models with a common transfer integral for simplicity. The total number of electrons is set the same as the number of sites. 
\begin{eqnarray}
H & = & \sum_i (\epsilon_i + v_i) n_i 
+ \sum_i U_i (n_{i\uparrow}-1/2)(n_{i\downarrow}-1/2)
\nonumber \\ & 
- & \sum_{i,\sigma} \left[ 
t_{i,i+1}(t) c^\dagger_{i,\sigma} c_{i+1,\sigma} +
t_{i+1,i}(t) c^\dagger_{i+1,\sigma} c_{i,\sigma} 
\right] 
\;,
\end{eqnarray}
where $ c^\dagger_{i,\sigma} $ ($c_{i,\sigma} $) creates (annihilates) an electron with spin $ \sigma $ at site $ i $, $ n_{i\sigma} = c^\dagger_{i,\sigma} c_{i,\sigma} $, and $ n_i = \sum_\sigma n_{i\sigma} $. The site energy $ \epsilon_i $ is set at $ \phi_\mathrm{L} $ in the left electrode at $ 1 \leq i \leq (L_\mathrm{e}-1)/2 $, at 0 in the insulator at $ (L_\mathrm{e}+1)/2 \leq i \leq L-(L_\mathrm{e}+1)/2 $, and at $ \phi_\mathrm{R} $ in the right electrode at $ L-(L_\mathrm{e}-1)/2 \leq i \leq L $, where $ L $ is the total number of sites and $ L_\mathrm{e} $ the number of sites in the electrodes (Fig.~\ref{fig:geometry}). 
The absolute value of the transfer integral $ \mid t_{i,i+1}(t) \mid $ is set at $ t_\mathrm{c} $ if either $ i $ or $ i+1 $ is in the insulator and at $ t_\mathrm{e} $ otherwise. When we consider a Mott insulator, the on-site repulsion $ U_i $ is set at $ U $ in the insulator and at 0 in the electrodes. For a band insulator, $ U $=0 and the transfer integral $ t_\mathrm{c} $ above is replaced by $ t_\mathrm{c} -(-1)^i \delta t $. 

A scalar potential $ v_i $ is introduced above to account for redistribution of electrons at interfaces to form barriers, compensating work-function differences $ \phi_\mathrm{L} $ and $ \phi_\mathrm{R} $ in equilibrium. The periodic boundary condition is imposed on $ v_i $: $ v_0 = v_L $. The applied voltage $ V $ is so defined that it is positive when the right electrode has a lower potential (for electrons) than the left and the current (without multiplication of charge) flows to the right. In order to maintain the periodic boundary condition for finite $ V $, we introduce the Peierls phase into the transfer integral, 
\begin{equation}
t_{i,i+1}(t) = t^\ast_{i+1,i}(t) = \mid t_{i,i+1}(t) \mid \exp 
\left[ -i\frac{ea}{\hbar} \int^t dt' E(t') \right]
\;,
\end{equation}
where $ t $ denotes time, $ e $ the absolute value of the electronic charge, $ a $ the lattice constant, and $ E(t) $ the averaged electric field defined by $ E(t) = -V/(La) $. \cite{corr_yonemitsu_jpsj05} 

By adding the vector potential introduced into the Peierls phase to the scalar potential $ v_i $, the total potential $ \psi_i $ is given by $ \psi_i = v_i - V (i/L -1/2) $. Although the potential $ \psi_i $ is defined on lattice points, we solve the Poisson equation in the continuum space, 
\begin{equation}
\frac{d^2 \psi}{d x^2} = -V_\mathrm{P} \left( \langle n \rangle - 1 \right) 
\;\mathrm{for}\; L_\mathrm{e}/2 < x < L - L_\mathrm{e}/2 
\;,
\label{eq:poisson}
\end{equation}
and $ d^2 \psi/d x^2 $=0 otherwise, where the potential $ \psi $ and the expectation value of the electron density per site $ \langle n \rangle $ are functions of $ x $, and $ V_\mathrm{P} $ comes from the long-range Coulomb interaction. In order to match the Fermi levels, we set the boundary condition, i.e., the potentials in the metallic electrodes, as 
\begin{eqnarray}
\psi(x) & = & 
-\phi_\mathrm{L} + V/2 \;\;\;\mathrm{for}\; 0 < x < L_\mathrm{e}/2 \;, \nonumber \\ 
\psi(x) & = & 
-\phi_\mathrm{R} - V/2 \;\;\;\mathrm{for}\; L - L_\mathrm{e}/2 < x < L \;.
\label{eq:boundary}
\end{eqnarray}
In order to solve the Poisson equation analytically, we introduce a simplified density-potential relation, which is known to work well for the potential distribution even in strongly correlated electron systems, \cite{oka_prl05} 
\begin{equation}
-\frac{dn(\psi)}{d\psi} = \kappa \equiv \frac{2}{W-\Delta}
\;, \label{eq:density_pot}
\end{equation}
with a constant compressibility $ \kappa $, the bandwidth $ W $, and the gap $ \Delta $. We have confirmed, by using the expectation value $ \langle n \rangle $ with respect to the wave function obtained from the time-dependent Schr\"odinger equation and by solving the Poisson equation for $ \psi $ simultaneously, that the current-voltage characteristics are qualitatively unchanged (Compare the present results with those in Ref.~\onlinecite{yonemitsu_pacifichem}). For the expression of $ \psi(x) $, see Appendix. 

First, we set $ V $ at zero and obtain the ground state either exactly (for small systems with $ L \leq $14) or in the unrestricted Hartree-Fock approximation. Then, we set $ V $ at a finite value to solve the time-dependent Schr\"odinger equation. For exact many-electron wave functions, the exponential evolution operator with time slice $ dt $=10$^{-2}$ is expanded to the 15th order. For self-consistent Hartree-Fock wave functions, the evolution operator with time slice $ dt $=10$^{-4}$ is decomposed with the help of the Suzuki-Trotter formula so as to be accurate to the order of $ dt^2 $. The current density $ J(t) $ is averaged over the period, $ 0 < t < \Delta t $ with $ \Delta t = 2 \pi \hbar L /(4eV) $, \cite{corr_yonemitsu_jpsj05} to give $ I $, where $ J(t) $ is given by \cite{oka_prl03,yonemitsu_jpsj05,corr_yonemitsu_jpsj05}
\begin{equation}
J(t) = (1/L)\sum_{i,\sigma} \left[
i t_{i+1,i}(t) c^\dagger_{i+1,\sigma} c_{i,\sigma}
- i t_{i,i+1}(t) c^\dagger_{i,\sigma} c_{i+1,\sigma}
\right]
\;.
\end{equation}

\section{Numerical Results \label{sec:theory}}

\begin{figure}
\includegraphics[height=6cm]{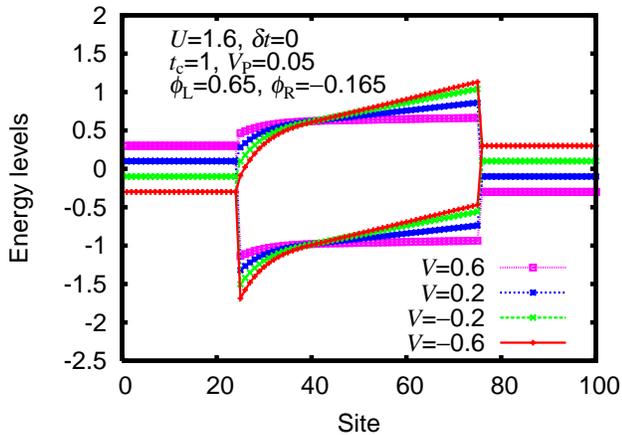}
\caption{(Color online) Energy levels, $ \epsilon_i + \psi_i \pm U_i/2 $, where $ \psi_i $ is the solution to the Poisson equation (\ref{eq:poisson}) with the simplified density-potential relation (\ref{eq:density_pot}) for $ L $=100, $ L_\mathrm{e} $=49, $ W $=4.0, $ \Delta $=0.33 (corresponding to $ U $=1.6, $ \delta t $=0 or $ U $=0, $ \delta t $=0.0825 with $ t_\mathrm{c} $=1), $ V_\mathrm{P} $=0.05, $ \phi_\mathrm{L} $=0.65, and $ \phi_\mathrm{R} $= $-$0.165. A Schottky barrier is formed at the left interface, and a positive (negative) $ V $ for right- (left-)going electrons corresponds to a forward (reverse) voltage lowering (increasing) the barrier height. 
\label{fig:barrier}}
\end{figure}

The analytic solution $ \psi_i $ given in Appendix plus the site energy $ \epsilon_i - U_i/2 $ is plotted in Fig.~\ref{fig:barrier}. The energy levels for doubly occupied sites, $ \epsilon_i + \psi_i + U_i/2 $ in the case of a Mott insulator, are also plotted as a guide. This result is used later in calculating the time evolution of the system. Here, the work function of the right electrode is set to be matched with the top of the lower Hubbard or valence band. The work function of the left electrode is set to be much lower than that of the insulator or that of the right electrode. A Schottky barrier is formed at the left interface. Its barrier height is lowered (increased) for right- (left-)going electrons, so that a positive (negative) $ V $ is a forward (reverse) voltage. 

\begin{figure}
\includegraphics[height=12cm]{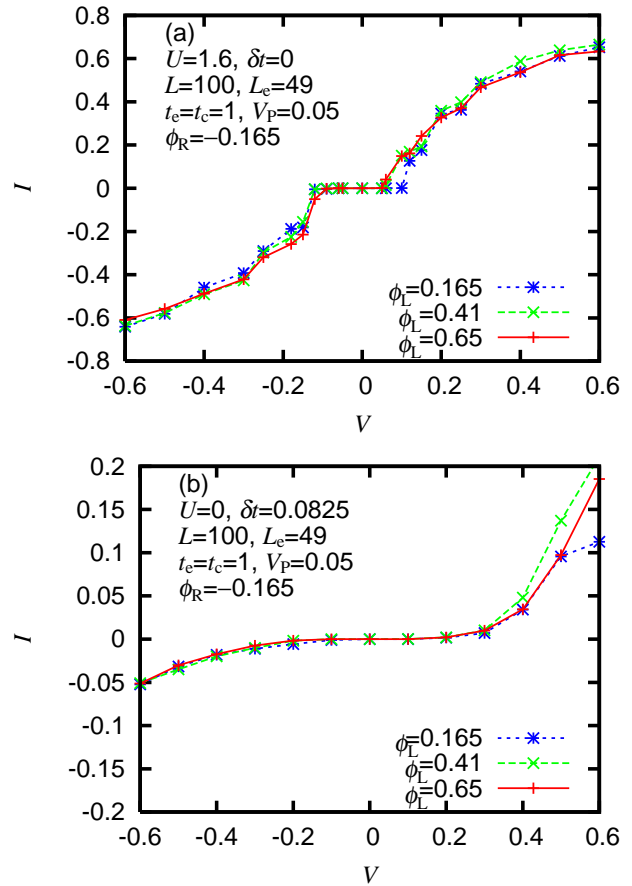}
\caption{(Color online) Current-voltage characteristics for (a) Mott insulator with $ U $=1.6, $ \delta t $=0, and (b) band insulator with $ U $=0, $ \delta t $=0.0825, obtained by the time-dependent Hartree-Fock approximation for $ L $=100 and $ L_\mathrm{e} $=49. Other parameters are $ t_\mathrm{e} $=$ t_\mathrm{c} $=1, $ V_\mathrm{P} $=0.05, $ \phi_\mathrm{R} $= $-$0.165, with different $ \phi_\mathrm{L} $ values as indicated. Rectification that is clearly seen for the band insulator is suppressed for the Mott insulator. 
\label{fig:Mott_vs_band}}
\end{figure}

In this situation, one usually expects that the absolute value of the current density $ I $, $ \mid I \mid $, is large (small) for $ V > $0 ($ V < $0), i.e., rectification behavior. \cite{sze-ng_book07} It is indeed reproduced for a band insulator, as shown in Fig.~\ref{fig:Mott_vs_band}(b). However, the current-voltage characteristics for a Mott insulator are qualitatively different, as shown in Fig.~\ref{fig:Mott_vs_band}(a). 
Namely, the absolute value of the current density is insensitive to the sign of the voltage, so that the current density $ I $ is almost an odd function of $ V $ for a wide range of $ \phi_\mathrm{L} $. The rectifying action is strongly suppressed for the Mott insulator. This property is maintained even for large $ \phi_\mathrm{L} $. We have performed numerical calculations with different sizes $ L $, $ L_\mathrm{e} $ and with different Coulomb parameters $ V_\mathrm{P} $ to find that this property is quite robust. 

\begin{figure}
\includegraphics[height=12cm]{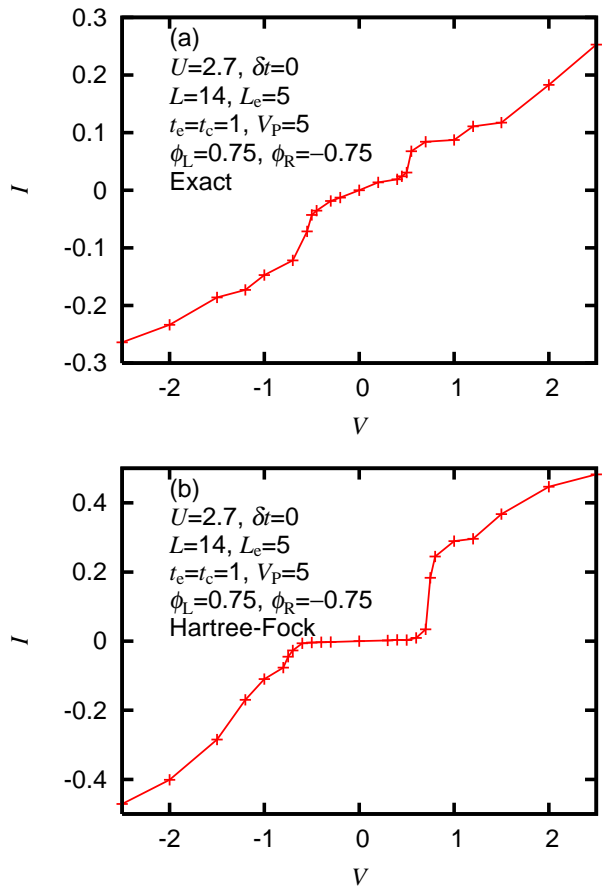}
\caption{(Color online) Current-voltage characteristics for (a) exact many-electron wave function, and (b) Hartree-Fock wave function, on a small system with $ L $=14 and $ L_\mathrm{e} $=5. Other parameters are $ U $=2.7, $ \delta t $=0, $ t_\mathrm{e} $=$ t_\mathrm{c} $=1, $ V_\mathrm{P} $=5, $ \phi_\mathrm{L} $=0.75, and $ \phi_\mathrm{R} $= $-$0.75. The overall characteristics showing suppressed rectification are obtained for both wave functions. 
\label{fig:exact_vs_HF}}
\end{figure}

In order to confirm that this property is not an artifact of the approximation, we then use exact many-electron wave functions on small systems and compare their current-voltage characteristics with those by the time-dependent Hartree-Fock approximation in Fig.~\ref{fig:exact_vs_HF}. 
Here a large Coulomb parameter $ V_\mathrm{P} $ is used to accommodate the band bending within the small insulator of size $ L- L_\mathrm{e} $. We also use $ U $=0 and $ \delta t $=0.375 for a band insulator with gap $ \Delta $=1.5, which is close to the gap in Fig.~\ref{fig:exact_vs_HF}(b), to confirm that rectification survives in the band insulator on this small system. In contrast, rectification turns out to be suppressed in the Mott insulator irrespective of whether exact many-electron wave functions are used or the time-dependent Hartree-Fock approximation is employed, although they show a quantitative difference. 

The similar qualitative difference between Mott and band insulators is demonstrated by field-effect carrier injections. \cite{hasegawa_prb04,yonemitsu_jpsj05} The field-effect characteristics are always ambipolar for Mott insulators and unipolar for band insulators when the work functions are different between the insulator and the metallic electrode. This field-effect property (with drain current $ I_\mathrm{D} $, drain voltage $ V_\mathrm{D} $, and gate voltage $ U_\mathrm{G} $ in the notations of Ref.~\onlinecite{yonemitsu_jpsj05}) can be shown to be closely related to the present interfacial property. The present result of $ I $ as a function of $ V $ is denoted by $ I = f(V) $. Then, in the geometry of Ref.~\onlinecite{yonemitsu_jpsj05}, the left (right) interface tends to produce a forward (backward) current density $ I_1 $ ($ I_2 $) by the potential difference $ U_\mathrm{G} + V_\mathrm{D}/2 $ ($ U_\mathrm{G} - V_\mathrm{D}/2 $). The total current is approximately given by $ I_\mathrm{D} \propto I_1 - I_2 = f(U_\mathrm{G} + V_\mathrm{D}/2)-f(U_\mathrm{G} - V_\mathrm{D}/2) \simeq V_\mathrm{D} f'(U_\mathrm{G}) $ for small $ V_\mathrm{D} $, where $ f'(V) $ is the derivative of $ f(V) $. Therefore, the suppressed rectification [$ f(V) $: odd function of $ V $] leads to ambipolar field-effect characteristics [$ f'(U_\mathrm{G}) $: even function of $ U_\mathrm{G} $]. The above relation between the interfacial property and the field-effect property through differentiation approximately holds for band insulators also. 

For the field-effect characteristics, the gate-bias polarity possessing the higher Schottky barrier at the metal-Mott-insulator interface is accompanied by more deviation from half filling (i.e., weakened umklapp scattering) inside the channel than that for the opposite gate-bias polarity. \cite{yonemitsu_jpsj05} Such a counterbalance is observed in detail and found in a very wide parameter space spanned by the work-function difference, the bandwidth difference, and the Coulomb parameter in the Poisson equation. In the present case, a very similar mechanism works: the bias-polarity dependence of the present charge-density distribution around the Schottky barrier is quite similar to the corresponding one of the field-effect transistor. Charge transport through a Mott insulator is not simply governed by the interface but determined as a whole including the interfacial and bulk regions. It is therefore much less sensitive to the details of the interfacial barrier potential than charge transport through a band insulator. Whether the voltage is forward or reverse is insignificant at metal-Mott-insulator interfaces. Such collective charge transport turns out to be realized by balancing the barrier effect with the correlation effect, as shown in Fig.~9 of Ref.~\onlinecite{yonemitsu_jpsj05}. 

\section{Experimental Results \label{sec:experiment}}

In the experiment, we prepared single crystals of two kinds of quasi-one-dimensional organic charge-transfer complexes; (BEDT-TTF)(F$_2$TCNQ) and K-TCNQ (TCNQ=tetracyanoquinodimethane). The latter is known as a Peierls-type (band) insulator composed of segregated stacks of TCNQ anion radicals. The stack of the compound is strongly dimerized along the stacking axis at room temperature. \cite{torrance_ssc93} The former is a Mott insulator composed of {\em side-by-side} arrangement of BEDT-TTF cation radicals. The molecular arrangement is free of dimerization down to low temperature ($ \sim $ 4 K). We fabricated asymmetric contacts on top of the single crystals by vacuum deposition with a gap of about 100 $ \mu $m, to measure the current along the quasi-one-dimensional stacks or chains. Cathodes are fabricated with 100 nm of magnesium (work-function $ \phi $= 3.66 eV), which is coated with thin (3 nm) silver layers to prevent the oxidation of magnesium. The contacts form Schottky junctions with both crystals. In contrast, anodes are fabricated with 50 nm of gold ($ \phi $= 5.1 eV) for (BEDT-TTF)(F$_2$TCNQ) crystals and 50 nm of silver ($ \phi$= 4.26 eV) for K-TCNQ crystals. The contacts form ohmic junctions with the respective crystals at room temperature. The DC current-voltage characteristics were measured with the use of the semiconductor parameter analyzer (Agilent E5270).

\begin{figure}
\includegraphics[height=14cm]{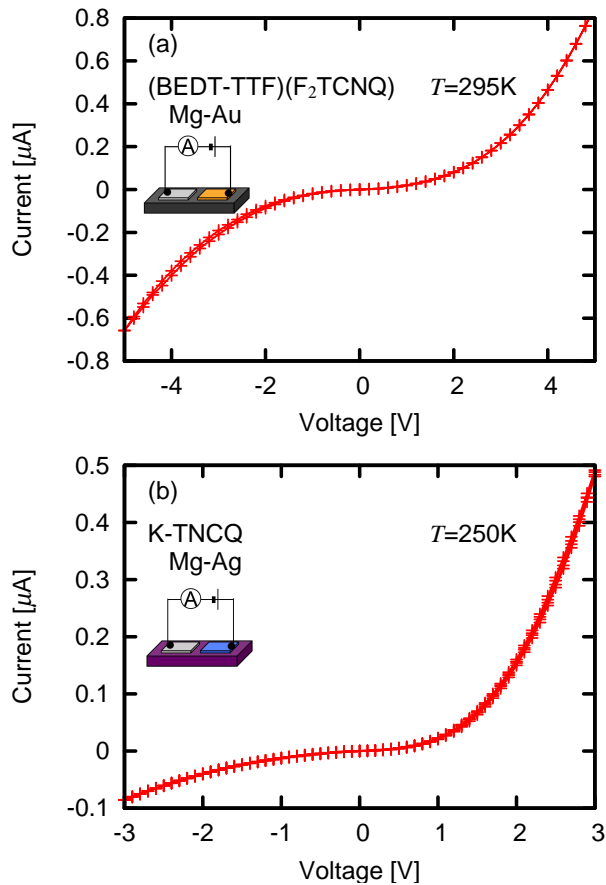}
\caption{(Color online) (a) Current-voltage characteristics of (BEDT-TTF)(F$_2$TCNQ) with magnesium cathode and gold anode at 295K. (b) Current-voltage characteristics of K-TCNQ with magnesium cathode and silver anode at 250K. 
\label{fig:IV_Mott_vs_band}}
\end{figure}

Current-voltage characteristics of the devices are shown in Fig.~\ref{fig:IV_Mott_vs_band}. It is found that the device with K-TCNQ exhibits rectifying nature with the rectification ratio of about 5 [Fig.~\ref{fig:IV_Mott_vs_band}(b)]. The results clearly show the formation of typical Schottky junctions in K-TCNQ with magnesium. In sharp contrast, the device with (BEDT-TTF)(F$_2$TCNQ) does not show rectification, while it exhibits distinct nonlinear features in the low-voltage range at both polarities [Fig.~\ref{fig:IV_Mott_vs_band}(a)]. The observed nonlinearity should be ascribed to the cathode characteristics with magnesium, since the device with gold contacts for both the cathode and the anode show much more conductive characteristics. It should be noted here that electron correlations are actually strong in K-TCNQ, but the dimerization-induced backward scattering does not allow the balance between the barrier effect and the correlation effect, leading to the band-insulator-like behavior. Thus, we found that these experimental results are consistent with the theoretical arguments presented above.

\section{Summary \label{sec:summary}}
In order to study the effect of electron correlation on charge transport through metal-insulator interfaces, we consider a one-dimensional Mott insulator, to which two metallic electrodes with different work functions are attached, and observe its current-voltage characteristics. The spatial dependence of the potential including Schottky barriers at interfaces is theoretically modeled by an analytic solution to the Poisson equation with a simplified density-potential relation. The Hubbard model is used for a Mott insulator and the tight-binding model with alternating transfer integrals for a band insulator. For small systems, we have compared numerical results obtained through the time-dependent Hartree-Fock approximation with those of exact many-electron wave functions. The qualitative characteristics regarding the current-voltage asymmetry are not lost by this approximation, which is used for large systems. Rectification is shown to be strongly suppressed at metal-Mott-insulator interfaces even with large work-function differences. This property dominated by an interface is shown to be closely related to the previously explained, ambipolar field-effect characteristics of a one-dimensional Mott insulator with two interfaces. This collective charge transport through metal-Mott-insulator interfaces is a consequence of the fact that backward scatterings at interfaces and umklapp scatterings inside the Mott insulator are balanced. The suppression of rectification is experimentally observed in the current-voltage characteristics of the quasi-one-dimensional Mott insulator (BEDT-TTF)(F$_2$TCNQ), with magnesium cathode forming a Schottky junction and gold anode forming an ohmic junction, at room temperature.

\begin{acknowledgments}
This work was supported by Grants-in-Aid and the Next Generation SuperComputing Project (Nanoscience Program) from the Ministry of Education, Culture, Sports, Science and Technology, Japan.
\end{acknowledgments}

\appendix*
\section{Potential $ \psi(x) $}
We adopt the simplified density-potential relation (\ref{eq:density_pot}), which implies 
\begin{eqnarray}
n(\psi) & = & 
2 \;\;\;\;\;\;\;\;\;\;\;\;\;\;\;\;\;\;\;\;\;\;\;\;\;
\mathrm{for}\; -\infty < \psi < -W/2 \;, \nonumber \\
n(\psi) & = & 
1-\kappa \left( \psi + \Delta/2 \right) 
\;\mathrm{for}\; -W/2 < \psi < -\Delta/2 \;, \nonumber \\
n(\psi) & = & 
1 \;\;\;\;\;\;\;\;\;\;\;\;\;\;\;\;\;\;\;\;\;\;\;\;\;
\mathrm{for}\; -\Delta/2 < \psi < \Delta/2 \;, \nonumber \\
n(\psi) & = & 
1-\kappa \left( \psi - \Delta/2 \right) 
\;\mathrm{for}\; \Delta/2 < \psi < W/2 \;, \nonumber \\
n(\psi) & = & 
0 \;\;\;\;\;\;\;\;\;\;\;\;\;\;\;\;\;\;\;\;\;\;\;\;\;
\mathrm{for}\; W/2 < \psi < \infty \;. 
\end{eqnarray}
Then, the solution to the Poisson equation (\ref{eq:poisson}) is given by 
\begin{equation}
\psi(x) = -\Delta/2 \pm c_0 
\sinh \left[ \sqrt{ \kappa V_\mathrm{P} } (x-x_0) \right] \mathrm{for}\; 1 < n < 2 
\end{equation}
with sign $ + $ for $ x < x_0 $ and $ - $ for $ x_0 < x $, and 
\begin{equation}
\psi(x) = \Delta/2 \pm c_0 
\sinh \left[ \sqrt{ \kappa V_\mathrm{P} } (x-x_0) \right] \mathrm{for}\; 0 < n < 1 
\end{equation}
with sign $ + $ for $ x_0 < x $ and $ - $ for $ x < x_0 $, where $ c_0 $ and $ x_0 $ are constants. Taking the boundary condition (\ref{eq:boundary}) into account, we obtain (with definitions $ \phi_\mathrm{L}(V) \equiv \phi_\mathrm{L} - V/2 $ 
and $ \phi_\mathrm{R}(V) \equiv \phi_\mathrm{R} + V/2 $) for example: 
\begin{eqnarray}
\mathrm{i)} & & \mathrm{if} 
-\frac{\Delta}{2} \leq -\phi_\mathrm{L}(V) \leq \frac{\Delta}{2} 
\;\mathrm{and} 
-\frac{\Delta}{2} \leq -\phi_\mathrm{R}(V) \leq \frac{\Delta}{2} 
\;, \nonumber \\
& & \psi(x) = \frac{
\phi_\mathrm{L}(V)(x-L+L_\mathrm{e}/2) 
- \phi_\mathrm{R}(V)(x-L_\mathrm{e}/2) }{ L- L_\mathrm{e} } \nonumber \\
& & \;\mathrm{for}\; L_\mathrm{e}/2 < x < L - L_\mathrm{e}/2 \;;
\end{eqnarray}
\begin{eqnarray}
\mathrm{ii)} & & \mathrm{if} 
-\frac{W}{2} \leq -\phi_\mathrm{L}(V) < -\frac{\Delta}{2} 
\;\mathrm{and} 
-\frac{\Delta}{2} \leq -\phi_\mathrm{R}(V) \leq \frac{\Delta}{2} 
\;, \nonumber \\
& & \psi(x) = 
-\Delta/2 + c_0 
\sinh \left[ \sqrt{ \kappa V_\mathrm{P} } (x-x_0) \right] \nonumber \\
& & \;\mathrm{for}\; L_\mathrm{e}/2 < x < x_0 \;, \nonumber \\
& & \psi(x) = 
\frac{ (\Delta/2)(x-L+L_\mathrm{e}/2)
-\phi_\mathrm{R}(V)(x-x_0) }{ L-L_\mathrm{e}/2-x_0 } \nonumber \\
& & \;\mathrm{for}\; x_0 < x < L-L_\mathrm{e}/2 \;,
\end{eqnarray}
where $ x_0 $ is the solution to 
\begin{equation}
\frac{\sinh
\left[ \sqrt{ \kappa V_\mathrm{P} }(x_0-L_\mathrm{e}/2) \right]}
{\sqrt{ \kappa V_\mathrm{P} }(L-L_\mathrm{e}/2-x_0)}=
\frac{-\Delta/2+\phi_\mathrm{L}(V)}
{- \phi_\mathrm{R}(V)+\Delta/2}
\;,
\end{equation}
and $ c_0 $ is given by 
\begin{equation}
c_0 = \frac{- \phi_\mathrm{R}(V)+\Delta/2}
{\sqrt{ \kappa V_\mathrm{P} }( L-L_\mathrm{e}/2-x_0)}
\;;
\end{equation}
\begin{eqnarray}
\mathrm{iii)} & & \mathrm{if} 
-\frac{\Delta}{2} \leq -\phi_\mathrm{L}(V) \leq \frac{\Delta}{2} 
\;\mathrm{and} 
-\frac{W}{2} \leq -\phi_\mathrm{R}(V) < -\frac{\Delta}{2} 
\;, \nonumber \\
& & \psi(x) = 
\frac{ -(\Delta/2)(x-L_\mathrm{e}/2)
+\phi_\mathrm{L}(V)(x-x_0) }{ x_0-L_\mathrm{e}/2 } \nonumber \\
& & \;\mathrm{for}\; L_\mathrm{e}/2 < x < x_0 \;, \nonumber \\
& & \psi(x) = 
-\Delta/2 - c_0 
\sinh \left[ \sqrt{ \kappa V_\mathrm{P} } (x-x_0) \right] \nonumber \\
& & \;\mathrm{for}\; x_0 < x < L-L_\mathrm{e}/2 \;,
\end{eqnarray}
where $ x_0 $ is the solution to 
\begin{equation}
\frac{\sqrt{ \kappa V_\mathrm{P} }(x_0-L_\mathrm{e}/2)}
{\sinh
\left[\sqrt{ \kappa V_\mathrm{P} }(L-L_\mathrm{e}/2-x_0) \right]}=
\frac{-\phi_\mathrm{L}(V)+\Delta/2}
{-\Delta/2+\phi_\mathrm{R}(V)}
\;,
\end{equation}
and $ c_0 $ is given by 
\begin{equation}
c_0 = \frac{-\phi_\mathrm{L}(V)+\Delta/2}
{\sqrt{ \kappa V_\mathrm{P} }(x_0-L_\mathrm{e}/2)}
\;;
\end{equation}
\begin{eqnarray}
\mathrm{iv)} & & \mathrm{if} 
-\frac{W}{2} \leq -\phi_\mathrm{L}(V) < -\frac{\Delta}{2} 
\;\mathrm{and}\; 
\frac{\Delta}{2} < -\phi_\mathrm{R}(V) \leq \frac{W}{2}
\;, \nonumber \\
& & \psi(x) = 
-\Delta/2 + c_0 
\sinh \left[ \sqrt{ \kappa V_\mathrm{P} } (x-x_0) \right] \nonumber \\
& & \;\mathrm{for}\; L_\mathrm{e}/2 < x < x_0 \;, \nonumber \\
& & \psi(x) = 
\frac{\Delta}{x_1-x_0}\left( x - \frac{x_1+x_0}{2} \right) 
\;\mathrm{for}\; x_0 < x < x_1 \;, \nonumber \\
& & \psi(x) = 
\Delta/2 + c_0 
\sinh \left[ \sqrt{ \kappa V_\mathrm{P} } (x-x_1) \right] \nonumber \\
& & \;\mathrm{for}\; x_1 < x < L-L_\mathrm{e}/2 \;,
\end{eqnarray}
where $ x_0 $ and $ x_1 $ are the solutions to 
\begin{equation}
\frac{\sinh
\left[ \sqrt{ \kappa V_\mathrm{P} }(x_0-L_\mathrm{e}/2) \right]}
{\sqrt{ \kappa V_\mathrm{P} }(x_1-x_0)}=
\frac{-\Delta/2+\phi_\mathrm{L}(V)}{\Delta}
\;,
\end{equation}
and 
\begin{equation}
\frac{\sqrt{ \kappa V_\mathrm{P} }(x_1-x_0)}
{\sinh
\left[\sqrt{ \kappa V_\mathrm{P} }(L-L_\mathrm{e}/2-x_1) \right]}=
\frac{\Delta}{- \phi_\mathrm{R}(V)-\Delta/2}
\;,
\end{equation}
and $ c_0 $ is given by 
\begin{equation}
c_0 = \frac{\Delta}{\sqrt{ \kappa V_\mathrm{P} }(x_1-x_0)}
\;.
\end{equation}

\bibliography{rect07_01}

\begin{thebibliography}{17}
\bibitem[{cor()}]{corr_yonemitsu_jpsj05}
\bibinfo{note}{In Ref.~\onlinecite{yonemitsu_jpsj05}, $ L $ in Eq.~(4) should
  read $ La $ with $ a $ being the lattice constant, the negative sign in front
  of $ \Sigma $ in Eq.~(10) should be omitted, and $ \Delta t = 2 \pi \hbar L
  /(4eV_\mathrm{D}) $ below Eq.~(10).}
\end{thebibliography}

\end{document}